\documentclass[aps,prb,amsmath,amssymb,reprint, showpacs, superscriptaddress]{revtex4-1}
\usepackage{graphicx}
\usepackage{epstopdf}
\usepackage{subfigure}
\usepackage{setspace}
\usepackage{textcomp}
\usepackage{placeins}
\usepackage{color}
\usepackage{xspace}
\usepackage{endnotes}
\usepackage{soul}
\usepackage{amsmath}

\usepackage{hyperref}

\begin{document}

\title{High-Resolution Hard X-ray Magnetic Imaging with Dichroic Ptychography}

\author{Claire~Donnelly}
\affiliation{Laboratory for Mesoscopic Systems, Department of Materials, ETH Zurich, 8093 Zurich, Switzerland}
\affiliation{Paul Scherrer Institute, 5232 Villigen PSI, Switzerland}
\author{Valerio~Scagnoli}
\email{valerios@ethz.ch}
\affiliation{Laboratory for Mesoscopic Systems, Department of Materials, ETH Zurich, 8093 Zurich, Switzerland}
\affiliation{Paul Scherrer Institute, 5232 Villigen PSI, Switzerland}
\author{Manuel~Guizar-Sicairos}
\affiliation{Paul Scherrer Institute, 5232 Villigen PSI, Switzerland}
\author{Mirko~Holler}
\affiliation{Paul Scherrer Institute, 5232 Villigen PSI, Switzerland}
\author{Fabrice~Wilhelm}
\affiliation{ESRF, 71 Avenue des Martyrs, 38000 Grenoble, France}
\author{Francois~Guillou}
\affiliation{ESRF, 71 Avenue des Martyrs, 38000 Grenoble, France}
\author{Andrei~Rogalev}
\affiliation{ESRF, 71 Avenue des Martyrs, 38000 Grenoble, France}
\author{Carsten~Detlefs}
\affiliation{ESRF, 71 Avenue des Martyrs, 38000 Grenoble, France}
\author{Andreas~Menzel}
\affiliation{Paul Scherrer Institute, 5232 Villigen PSI, Switzerland}
\author{J{\"o}rg~Raabe}
\affiliation{Paul Scherrer Institute, 5232 Villigen PSI, Switzerland}
\author{Laura~J.~Heyderman}
\affiliation{Laboratory for Mesoscopic Systems, Department of Materials, ETH Zurich, 8093 Zurich, Switzerland}
\affiliation{Paul Scherrer Institute, 5232 Villigen PSI, Switzerland}

\begin{abstract}
Imaging the magnetic structure of a material is essential to understanding the influence of the physical and chemical microstructure on its magnetic properties. Magnetic imaging techniques, however, have up to now been unable to probe three dimensional micrometer-sized systems with nanoscale resolution. Here we present the imaging of the magnetic domain configuration of a micrometer-thick FeGd multilayer with hard X-ray dichroic ptychography {at energies spanning both the Gd $L_3$ edge and the Fe $K$ edge, providing a high spatial resolution spectroscopic analysis of the complex X-ray magnetic circular dichroism.} With a spatial resolution reaching $45\,\rm{nm}$, this advance in hard X-ray magnetic imaging is the first step towards the investigation of buried magnetic structures and extended three-dimensional magnetic systems at the nanoscale. 
\end{abstract}

                            

\maketitle

For optimal integration of new magnetic materials and systems into technological applications, a detailed understanding of the magnetic structure and properties is required. In particular, for the next generation of nanocomposite permanent magnets, which have the potential to enable more efficient, smaller and lighter energy harvesters and motors \cite{Gutfleisch11}, an understanding of the influence of the physical and chemical microstructure on the magnetic configurations is of the utmost importance. X-ray imaging is an excellent candidate for such investigations \cite{Stohr_book,Stohr93}, offering high spatial resolution, element specificity, high sensitivity, and the ability to directly and non-destructively probe the magnetization of a material. However, so far, all sub-micrometer spatial resolution X-ray magnetic imaging has been performed with soft X-rays, with photon energies below $1.6\,\rm{keV}$, due to the strong magnetic signal available in this regime \cite{Eisebitt04,Chen90}. Whilst effective for probing the magnetism of thin films and nanostructures \cite{Eisebitt04,Fischer15}, soft X-rays are limited to systems with total material thickness below approximately $200\,\rm{nm}$ \cite{Streubel15,Fischer11}. {Neutron magnetic imaging allows the study of systems with thicknesses ranging from micrometers up to millimeters but only with a spatial resolution of tens to hundreds of micrometers\cite{Pfeiffer06,Betz16,Kardjilov08,Manke10}.} As a result, high spatial resolution magnetic investigations of thick permanent magnets to be used in microsystems such as electromagnetic sensors and energy harvesting devices \cite{Gibbs04,Niarchos03,Jiang11}, as well as the study of buried structures integrated into, for example, silicon-based devices \cite{Chappert07}, have not been possible up to now.

Higher energy X-rays offer a solution to such limitations, with recent advances providing nanometer spatial resolution imaging of structures with thickness on the order of many micrometers \cite{Holler14}. For magnetic imaging, however, hard X-rays have so far not been considered to be competitive. Indeed, soft X-rays, by utilizing the ${L_{2,3}}$ edges (2\textit{p}-3\textit{d}) of transition metals and the $M_{4,5}$ edges (3\textit{d}-4\textit{f}) of rare earths, directly probe the valence band of the magnetic material. In the hard X-ray regime, however, one indirectly probes the magnetization at the \textit{K} (1\textit{s}-4\textit{p}) edges of transition metals and the ${L_{2,3}}$ (2\textit{p}-5\textit{d}) edges of rare earths, resulting in a significantly weaker magnetic signal \cite{Hippert_book}. In addition, in previous studies of hard X-ray magnetic imaging a maximum spatial resolution of $2\,\mu\rm{m}$ has been reached for the imaging of two dimensional objects \cite{Lang01,Sato01,Takagaki06,Ueda10}. 

Here we present hard X-ray magnetic imaging at the nanoscale with a spatial resolution of $45\,\rm{nm}$. As a proof-of-principle, we image the magnetic domain configuration of a micrometer-thick FeGd film at energies spanning not only the Gd $L_3$ edge, with relatively strong magnetic contrast, but also the Fe $K$ edge. While the magnetic signal is substantially weaker at the Fe $K$ edge, we demonstrate that we can image the magnetic configuration of the Fe within the system, thus demonstrating the applicability of the technique to non-rare earth magnetic elements, which are of scientific interest as an important alternative to the scarce and expensive rare earth magnetic materials \cite{Kramer12}. {We perform a quantitative spectroscopic analysis of the full complex transmission function for energies spanning the Gd $L_3$ edge, providing a detailed insight into the relationship between the phase and absorption XMCD signals and confirming the quantitative nature of the technique.} Furthermore, we measure the magnetic configuration of a film with a thickness of $500\,\rm{nm}$, therefore demonstrating the potential of the technique to investigate thinner films. With hard X-ray magnetic imaging with nanoscale resolution, the investigation of magnetic configurations of a wide range of samples is now possible. 

In order to measure the weak magnetic signals in the hard X-ray regime, a high level of sensitivity is required, which is provided by ptychography. Ptychography is a scanning coherent diffractive imaging technique, which gives access to the full complex transmission function of an object \cite{Rodenburg07,Thibault08}. The sample is illuminated with a coherent beam, and far-field diffraction patterns are measured for many overlapping illumination positions. Both the object and illumination functions are then retrieved using an iterative reconstruction algorithm \cite{Thibault08,Guizar08,Maiden09}. When combined with three dimensional imaging techniques such as tomography, this access to the full complex transmission function provides a means to determine the three dimensional mapping of the electron \cite{Holler14,Diaz12} and atomic densities \cite{Donnelly15} within a sample. 

When applied to magnetic imaging, coherent diffractive imaging (CDI) techniques such as ptychography offer high resolution imaging with a number of key advantages \cite{Turner11,Tripathi11,Scherz07,Eisebitt04,Flewett12,Shi16}. For instance, the complex magnetic signal provided by both CDI and ptychography makes possible the imaging of ferromagnetic domains with linearly polarized light. Here the magnetic diffraction is isolated by subtracting the pure electronic scattering measured when the magnetic state is removed with a saturating magnetic field \cite{Turner11,Tripathi11}. In an investigation of the complex magnetic signal with soft X-ray holography it was reported that phase imaging before the absorption edge can minimize the radiation dose required \cite{Scherz07}. 


{
In this work we combine ptychography with X-ray magnetic circular dichroism (XMCD) in the hard X-ray regime. In particular we study FeGd multilayers, which are amorphous films with perpendicular magnetic anisotropy. XMCD is a resonant effect that occurs when the photon energy is tuned close to an atomic absorption edge. The electronic transitions between the resonant core level and the magnetically polarized valence band then become sensitive to the helicity of the X-rays and the projection of the sample's magnetization along the beam direction. Experiments are typically performed by reversing the X-ray helicity and/or the sample magnetization. A more in-depth description of XMCD is given by, for example, Ref. \onlinecite{Stohr_book}. In the case of ptychography, it is crucial to have a clear understanding of the complex index of refraction resulting from the interactions between the sample and the X-rays, especially when dichroic and birefringent effects cannot be neglected as in XMCD. In particular, the assumption in standard ptychography that the probe and object are complex functions that interact via a scalar product may not hold \cite{Ferrand15}. 
\begin{figure}
  \centering
\includegraphics{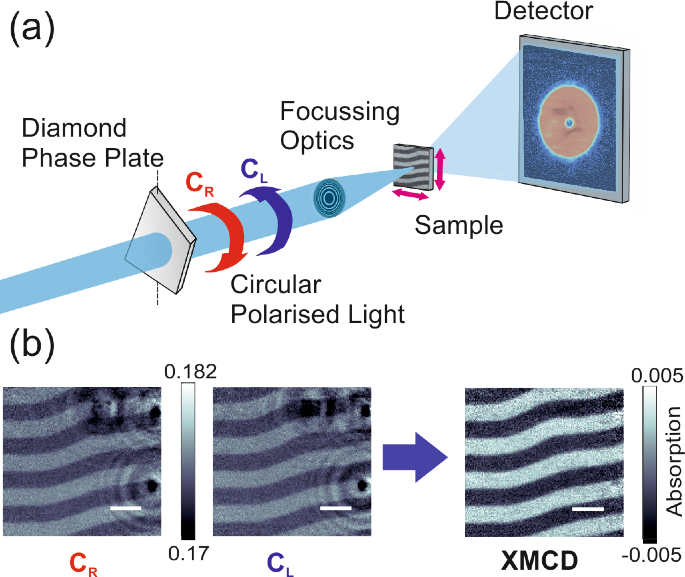}
	\caption{(a) The dichroic ptychography setup. A diamond phase plate converts linearly polarized light into $\rm{C_{L}}$ or $\rm{C_{R}}$ light, which is then focused close to the sample plane. A piezoelectric stage is used to scan the sample across the beam. (b) {The absorption part of reconstructed images} taken with $\rm{C_{L}}$ and $\rm{C_{R}}$ polarized light at the Gd $L_3$ edge with a photon energy of $7.2445\,\rm{keV}$, which contain both electron density and magnetic contributions. The difference of these images removes the electron density contrast of  Pt reference structures on the sample to give a purely magnetic image, i.e. with XMCD contrast. Scale bars represent $1\,\mu\rm{m}$.}
\label{fig:fig_setup}
\end{figure}

For the case of X-rays interacting with a magnetic material, one must consider both the electronic and magnetic contributions to the scattering factor\cite{Hannon88,Hill96,Lovesey96}:
\begin{equation}\label{eq:1}
\begin{split}
f= & f_c\big({\boldsymbol{ \epsilon}_f\mathbf{}^{\ast}\cdot \boldsymbol{\epsilon}}_i\big) - if_m^{(1)}\big({\boldsymbol{ \epsilon}_f^{\ast}\boldsymbol{\times \epsilon}_i\big)\cdot \mathbf{m(r)}} \\ &+ f_m^{(2)}\bigg(\boldsymbol{ \epsilon}_f^{ \ast}\cdot \mathbf{ m(r)}\bigg)\bigg(\boldsymbol{\epsilon}_i \cdot \mathbf{m(r)}\bigg)
\end{split}
\end{equation}
where $\mathbf{m}$ is the magnetization vector, $\boldsymbol{\epsilon}_i$ and $\boldsymbol{\epsilon}_f$ are the Jones polarization vectors of the incoming and scattered waves, respectively, $\ast$ denotes complex conjugation, and $f_c$,  $f_m^{(1)}$ and $f_m^{(2)}$ are the complex-valued charge, circular and linear dichroic magnetic scattering factors, respectively. For the case of small angle scattering, and under the approximation of perfectly circularly polarized incoming photons and a non-birefringent interaction, as is the case for the amorphous film studied in this work, Equation (\ref{eq:1}) is reduced to:
\begin{equation}\label{eq:2}
\begin{split}
f & = f_c - if_m^{(1)} \begin{pmatrix} \boldsymbol{\epsilon}_{C_{R}}^{\ast} \times \boldsymbol{\epsilon}_{C_{R}} & \boldsymbol{\epsilon}_{C_{L}}^{\ast} \times \boldsymbol{\epsilon}_{C_{R}} \\ \boldsymbol{\epsilon}_{C_{R}}^{\ast} \times \boldsymbol{\epsilon}_{C_{L}} & \boldsymbol{\epsilon}_{C_{L}}^{\ast} \times \boldsymbol{\epsilon}_{C_{L}} \end{pmatrix} \cdot \mathbf{m(r)}\\ & = f_c - if_m^{(1)} \begin{pmatrix} im_z & 0 \\ 0 & -im_z \end{pmatrix}
\end{split}
\end{equation}
For circularly right ($\rm{C_R}$) and left ($\rm{C_L}$) polarized incident light, there are therefore two non-zero matrix elements for the second term in Equation (2): the $\rm{C_L}$-to-$\rm{C_L}$ and the $\rm{C_R}$-to-$\rm{C_R}$ scattering channels, that are equal to $\pm f_m^{(1)} m_z$, where $m_z$ is the component of the magnetization along the propagation direction. The contribution of $f_m^{(2)}$  has been neglected as both scalar products are approximately zero for the case of circularly polarized light and perpendicular magnetic anisotropy, where the magnetization and polarization vectors are perpendicular. The latter term in Equation (\ref{eq:2}) is then the source of the XMCD signal that we probe with dichroic ptychography, and the result on the outgoing wave is a scalar addition to the scattering factor, which effectively results in a modification to the phase-shift and amplitude of the outgoing wave.

In considering a formalism for the ptychographic reconstructions, it is apparent that there are no polarization cross terms representing, for example, $\rm{C_R}$-to-$\rm{C_L}$ scattering channels between different polarizations. We can therefore conclude that for the case of perfectly circularly polarized incident light, the standard assumption that the probe and the object are complex functions that interact via a scalar product is valid. The interaction can be then expressed as in conventional ptychography as:
\begin{equation}
\mathbf{\Psi}_j(\mathbf{r}) = \mathbf{P(r)O(r-r}_j\mathbf{)}
\end{equation}
Where $\mathbf{\Psi}_j(\mathbf{r})$ is the exit field after the sample, $\mathbf{P(r)}$ is the incident illumination, and $\mathbf{O(r-r}_j$ is the transmissivity of the sample, which can be expressed in terms of the scattering factors as follows:
\begin{equation}
\mathbf{O(r-r}_j\mathbf{ )}=\int exp \bigg(i\frac{2\omega}{c}z\big[1-\frac{r_{\rm{e}}}{2\pi}\lambda^2n_{\rm{at}}f(\mathbf{r-r}_{\rm{j}})\big]\bigg)dz
\end{equation}
$\mathbf{\Psi}_j(\mathbf{r})$, $\mathbf{P(r)}$ and $\mathbf{O(r-r}_j\mathbf{ )}$ are all complex-valued functions of $\mathbf{r}$ which are the transverse Cartesian coordinates, and $j$ indicates the ptychography scan position.
}

Experimentally, we perform ptychography scans with $\rm{C_{L}}$ and $\rm{C_{R}}$ polarized light in order to non-destructively image magnetic features in our sample with both phase and absorption magnetic contrast, without the need to apply a saturating field. Explicitly, the complex transmission function is given by $T(E)=a(E)e^{i\phi(E)}$, where $a$ is the amplitude and $\phi$ the phase, which both depend strongly on the X-ray energy $E$ in the vicinity of an absorption edge. We define the dichroic signal as:
\begin{equation} 
A_{\rm{XMCD}} (\%) = \big(|a_{\rm{L}} |^{2}-|a_{\rm{R}} |^{2}\big)/\Delta_{\rm{edge}}   \times 100\,\% 		
\end{equation}
\begin{equation}
\phi_{\rm{XMCD}}=\phi_{\rm{L}}-\phi_{\rm{R}},
\end{equation}
where $A_{\rm{XMCD}}$ and $\phi_{\rm{XMCD}}$ are the absorption and phase XMCD signals, respectively, and $\Delta_{\rm{edge}}$ is the relative change in absorption across the absorption edge. $a_{\rm{L,R}}$ and $\phi_{\rm{L,R}}$ are the amplitude and phase of the complex transmission function, respectively, measured with left (L) and right (R) circularly polarized incident light.

\begin{figure*}
  \centering
\includegraphics[width=0.9\textwidth]{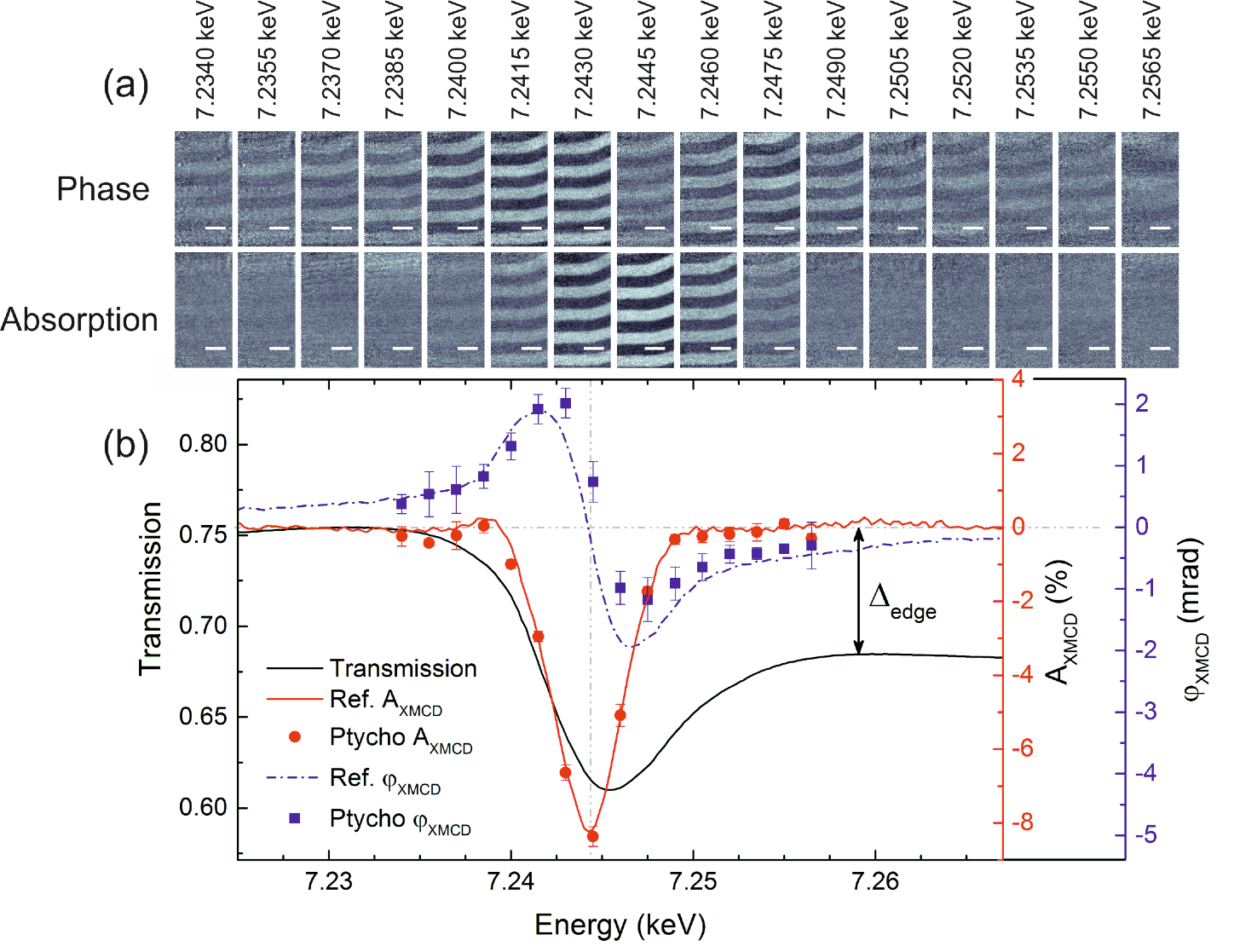}
	\caption{(a) XMCD ptychographic images of the magnetic domain configuration of the FeGd film obtained from the difference taken between images measured with $\rm{C_{R}}$ and $\rm{C_{L}}$ circular light. Phase (upper row) and absorption (lower row) contrast are given for a number of photon energies spanning the Gd $L_3$ edge. Scale bars represent $1\,\rm{\mu m}$. (b) The $A_{\rm{XMCD}}$ (red circles) and $\phi_{\rm{XMCD}}$ (blue squares) signal extracted from the images is presented, providing a direct comparison of the phase and absorption signals. The XMCD data are compared with a reference $A_{\rm{XMCD}}$ spectrum (red line) measured in fluorescence on a similar sample, and the corresponding $\phi_{\rm{XMCD}}$ reference spectrum calculated with the Kramers-Kronig relations (dashed blue line). A transmission spectrum of the sample (black line) is provided as a reference.}
\label{fig:fig2}
\end{figure*}

The dichroic ptychography setup (see Appendix \ref{setup} for more details) is shown in Fig.~\ref{fig:fig_setup}(a). In order to probe the XMCD, circularly polarized light was produced by converting linearly polarized light using a $500\,\rm{\mu m}$-thick diamond quarter-wave phase retarder \cite{Giles94} and the degree of polarization of the X-rays was determined using a polarization analyzer setup \cite{Scagnoli09}. The X-rays were found to be over $99\%$ circularly polarized, with the phase plate absorbing approximately $65\%$ of the incident light. 

Ptychography scans with a field of view of $5\times5\,\rm{\mu m}^2$ consisted of a diffraction pattern being measured at each of 278 scanning points placed on a Fermat spiral grid \cite{Huang14}. At each point, the diffraction pattern was measured with a 0.2 second exposure time using a single photon counting Pilatus 2M detector \cite{Henrich09,Kraft09} located $7.219(1)\,\rm{m}$ downstream of the sample. Each scan took approximately 112 seconds which includes the time taken for positioning the sample at each point with a piezo-electric motor. To enhance the signal-to-noise ratio, ten such scans were repeated for each energy, measured with left and right circularly polarized light. 

Ptychographic reconstructions with a field of view of $5\time 5\,\rm{\mu m}$ were performed using a gradient-based maximum likelihood estimation \cite{Guizar08b,Thibault12} with a maximum possible number of iterations of 500, although convergence was typically reached before 300 iterations. Further details on the the experimental setup and reconstruction are given in Appendix \ref{setup} and \ref{recons}.


Here we employ dichroic ptychography to image the magnetic configurations of $1\,\rm{\mu m}$ and $500\,\rm{nm}$ thick $\rm{Fe(0.41nm)/Gd(0.45nm)}$ multilayer films with perpendicular magnetic anisotropy. The multilayer films were deposited with RF magnetron sputtering on a $\rm{Si_3N_4}$ membrane with a growth rate of $0.15\,\rm{nm/s}$ at room temperature and characterized as described in Appendix \ref{materials_and_methods}.  To enable easier image alignment, platinum nanostructures were deposited on the film with focused electron beam deposition.

To demonstrate the isolation of the magnetic signal, the absorption part of the ptychography reconstructions taken with $\rm{C_{L}}$ and $\rm{C_{R}}$ polarized light at the Gd $L_3$ edge, with photon energy of $7.2445\,\rm{keV}$, are shown in Fig.~\ref{fig:fig_setup}(b). Along with the electron density contrast of the platinum alignment structures, magnetic stripe domains are visible and the magnetic contrast reverses when the polarization is changed from $\rm{C_{L}}$ to $\rm{C_{R}}$. The difference of the two images removes the electron density contrast, which in this case corresponds mainly to the platinum alignment structures, and results in an XMCD image with purely magnetic contrast, as shown in the rightmost image of Fig.~\ref{fig:fig_setup}(b).

The spatial resolution of a single absorption image was determined via Fourier ring correlation (FRC) \cite{Heel05} to be approximately $60\,\rm{nm}$. Averaging of five images resulted in a spatial resolution of $45\,\rm{nm}$, representing a fifty-fold increase in the achieved spatial resolution with dichroic ptychography compared with previous measurements with hard X-rays\cite{Lang01,Sato01,Takagaki06,Ueda10}. The spatial resolution of the phase images, however, was estimated to be $70-80\,\rm{nm}$. Thus, whilst the phase information typically offers better electron density contrast, in our experiment absorption contrast provided better magnetic contrast and spatial resolution. 
The edge sharpness of the domain walls was found to be approximately $60\,\rm{nm}$ by taking a line profile of the image intensity across a domain wall and measuring the $10\%-90\%$ edge sharpness. It should be noted that, thanks to recent advances in X-ray optics, which allow for the X-ray beam to be focused down to a $30\,\rm{nm}$ spot \cite{Mohacsi15}, a better resolution than that reported in Refs. \onlinecite{Lang01,Sato01,Takagaki06,Ueda10} would in principle be possible with other techniques such as scanning transmission X-ray microscopy. However, to the best of our knowledge such measurements have not been published.

To investigate the energy dependence of the complex XMCD signal and demonstrate the quantitative nature of the technique, dichroic ptychography was performed for different photon energies spanning the Gd $L_3$ edge, that ranged from $7.234\,\rm{keV}$ to $7.256\,\rm{keV}$ in steps of $1.5\,\rm{eV}$. Images of the phase and absorption components of the sample transmission are shown in Fig.~\ref{fig:fig2}(a). Below the edge, the magnetic phase contrast is inverted with respect to the absorption contrast, and can be seen to reverse when traversing the absorption edge. To measure the phase and absorption XMCD signal, a Gaussian distribution was fitted to the histograms of the XMCD contrast of the pixels within the domains. This fit is performed for a number of separate domains, and the mean and standard deviation of the XMCD signal calculated. They are given as a function of energy across the absorption edge in Fig.~\ref{fig:fig2}(b), represented by blue squares ($\phi_{\rm{XMCD}}$) and red circles ($A_{\rm{XMCD}}$), where the maximum absorption contrast is found to be $8.2\,\%$ of $\Delta_{\rm{edge}}$ at $7.2445\,\rm{keV}$, $1\,\rm{eV}$ below the absorption edge.

{
To determine the most effective energy at which to image, both the radiation dose and the magnetic contrast must be optimized. A single projection delivered an X-ray radiation dose of ${\rm 1.4\,MGy}$ and ${\rm 2.3\,MGy}$ on the sample just below the absorption edge at ${\rm 7.240\,keV}$ and at the absorption edge at ${\rm 7.2445\,keV}$, respectively. Although phase imaging before the edge may therefore result in reduced radiation damage, the on-resonance ${\rm A_{XMCD}}$ signal is approximately four times higher. Therefore, for the same image contrast, less cumulative dose is delivered when measuring in absorption on-resonance with a shorter exposure time. 
}

In order to confirm that the ptychography measurements provide a quantitative measurement of the complex XMCD signal, the ptychography $A_{\rm{XMCD}}$ spectrum was compared with a reference $A_{\rm{XMCD}}$ spectrum measured in fluorescence yield with a standard XMCD setup on an equivalent sample (see Appendix \ref{fluorescence} for more details) \cite{Rogalev01}, also plotted in Fig.~\ref{fig:fig2}(b) (red line). From the excellent agreement between the magnitude and form of the $A_{\rm{XMCD}}$ signals measured with ptychography and the fluorescence technique, it is evident that dichroic ptychography not only allows high quality imaging of the domain structure with unprecedented spatial resolution and sensitivity, but moreover provides a quantitative measure of the magnetic contrast. This allows the relative magnitude of vector components of the magnetization to be directly compared, which is critical for three dimensional magnetic investigations. 
The direct measurement of the $\phi_{\rm{XMCD}}$ signal (blue squares) is further compared with a $\phi_{\rm{XMCD}}$ spectrum calculated from reference absorption spectra using the Kramers-Kronig relations \cite{Hippert_book} (dashed blue line) in Fig.~\ref{fig:fig2}(b). While the calculated spectrum is symmetric about the inversion point, the measured $\phi_{\rm{XMCD}}$ data is shifted by $0.5\,\rm{eV}$, with the magnitude of its signal being asymmetric about the inversion point. Although not critical to the current work, these differences are likely to be associated with a physical effect that warrants further investigation in the future.

\begin{figure}
  \centering
\includegraphics{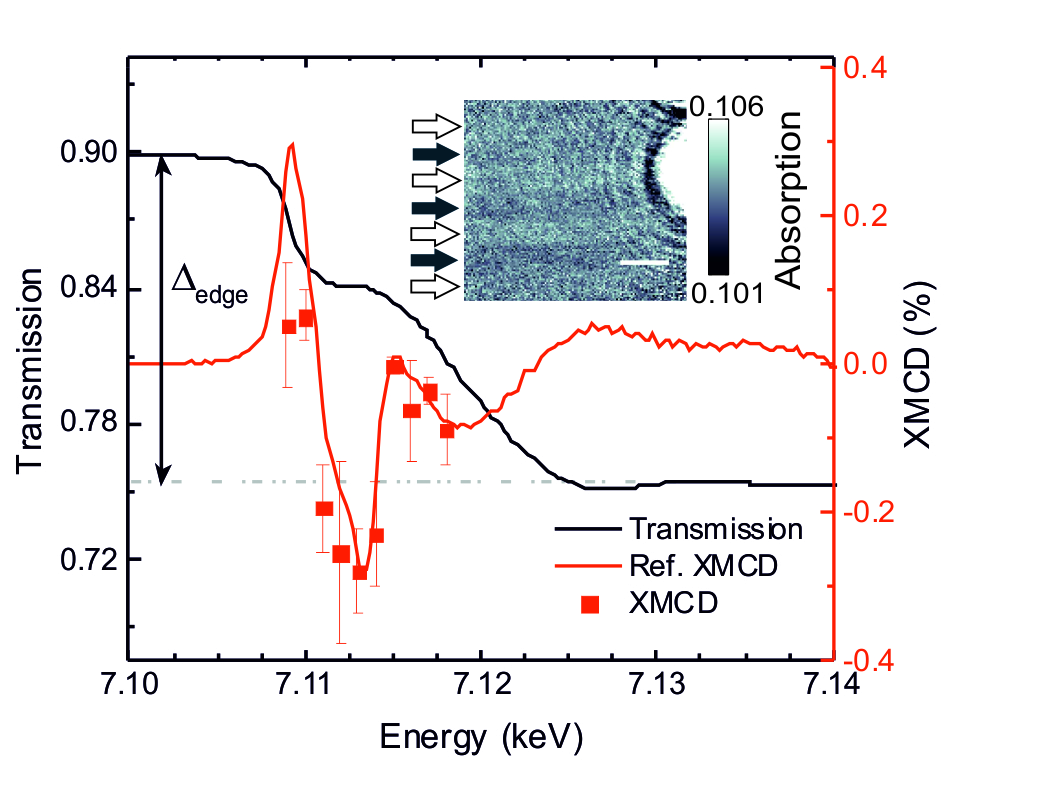}
	\caption{The $A_{\rm{XMCD}}$ signal measured with dichroic ptychography across the Fe $K$ edge (red squares) is plotted along with a reference XMCD spectrum (red line) measured in fluorescence on a similar film. A single polarization image is given in the inset, where the stripe domain pattern can be seen, along with the electron density contrast of the Pt markers. The domains are indicated by white and black arrows to the left of the image.}
\label{fig:fig3}
\end{figure}

{

The element specificity of X-rays enables the direct comparison of the magnetization of different elements within a sample. In the hard X-ray regime, the $K$ edges of transition metals are also available, offering the possibility of studying magnetic materials beyond rare earths.   
Although the magnetic signal at the Fe $K$ edge is an order of magnitude weaker than that found at the Gd $L_3$ edge \cite{Stohr_book}, we are able to image the magnetic domains at a range of energies across the Fe $K$ edge, demonstrating the sensitivity and quantitativeness of dichroic ptychography. The ptychography $A_{\rm{XMCD}}$ data extracted from the images measured at energies spanning the Fe $K$ edge are shown in Fig.~\ref{fig:fig3} (red squares). The spectrum is compared directly with a reference spectrum measured in fluorescence yield with a standard XMCD setup (red line), and a good level of agreement between the ptychography data and reference spectrum can be seen. 

To compare the magnetization of the Fe and Gd within the sample, the same section of the domain structure was imaged at the Fe $K$ edge and at the Gd $L_{\rm 3}$ edge. Magnetic absorption images measured with $C_{\rm L}$ polarized X-rays are shown in Fig. \ref{fig:supp_7}, and the magnetic domains are highlighted with black and white arrows to the left of the images. As the XMCD signal found at the Fe $K$ edge is approximately ten times weaker than that found at the Gd $L_{\rm 3}$ edge, the signal to noise ratio of the image measured at the Fe $K$ edge is visibly weaker. On comparing the magnetic contrast measured at the Gd $L_{\rm 3}$ edge (left image) and the Fe $K$ edge (right image), we observe the same domain structure that is reversed in contrast, indicating that the Fe and Gd are coupled antiferromagnetically, as expected for a ferrimagnet such as FeGd \cite{Tripathi11}.

\begin{figure}
  \centering
\includegraphics[width=0.4\textwidth]{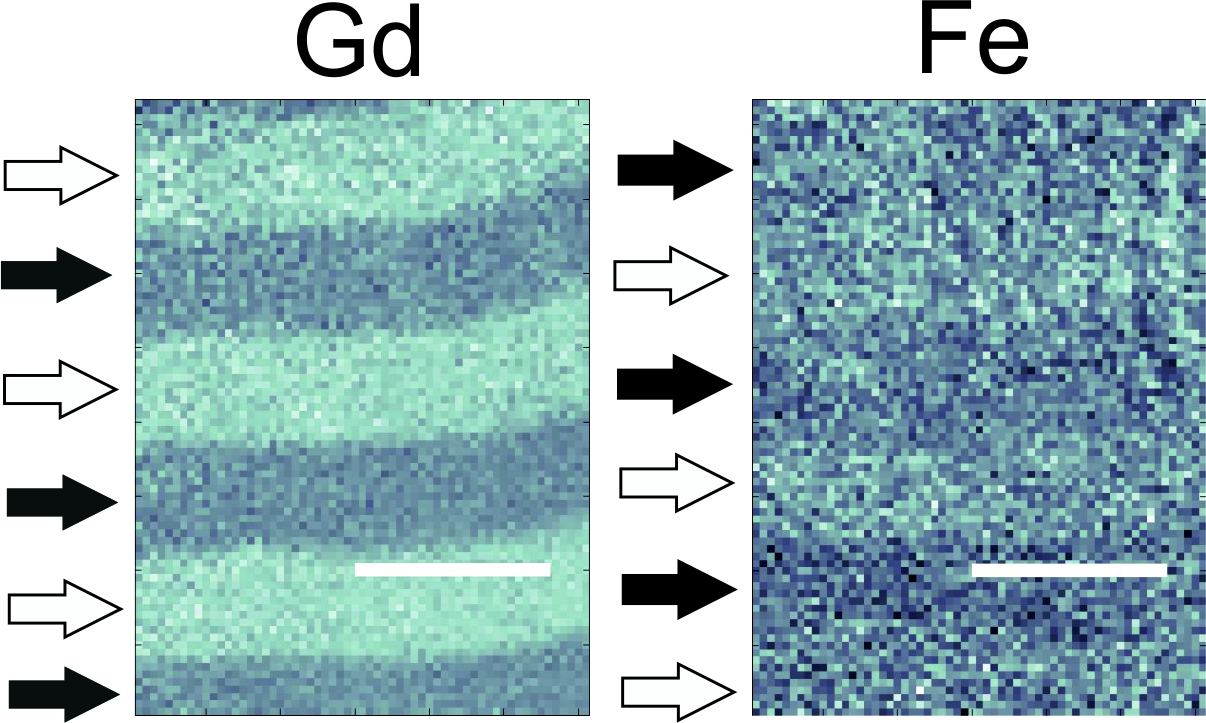}
	\caption{Absorption images of the domain structure of FeGd, measured with $C_L$ polarized light at the Gd $L_{\rm 3}$ edge with a photon energy of $7.2445\,\rm{keV}$ (left image) and the Fe $K$ edge with a photon energy of $7.113\,\rm{keV}$ (right image). Domains of opposite magnetization are indicated by black and white arrows on the left hand side of the images. The difference in contrast for the Fe $K$ edge and the Gd $L_3$ edge is an indication of the antiferromagnetic coupling between the Fe and Gd magnetic moments.  The scale bars represent $1\,{\rm \mu m}$.}
\label{fig:supp_7}
\end{figure}

 }

\begin{figure}
  \centering
\includegraphics[width=0.5\textwidth]{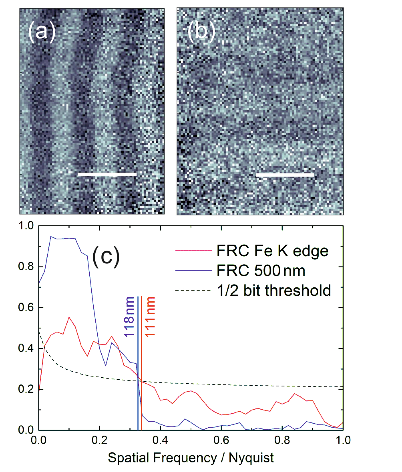}
	\caption{(a) XMCD absorption image of the magnetic domain structure of the $500\,\rm{nm}$ FeGd film measured at the Gd $L_{\rm{3}}$ edge. (b) Single polarization absorption image of the $1\,\rm{\mu m}$ film measured at the Fe $K$ edge. Scale bars represent $1\,\rm{\mu m}$. (c) The 1/2 bit criterion is used to provide an estimate of the spatial resolution of images \cite{Heel05}, which is found to be similar for the two images. }
\label{fig:spat_res}
\end{figure}

Finally, in order to investigate the suitability of hard X-ray ptychography for the imaging of thinner layers, the domain structure of a $500\,\rm{nm}$-thick FeGd film was imaged at the Gd $L_3$ edge with a photon energy of $7.243\,\rm{keV}$. The absorption image is shown in Figure \ref{fig:spat_res}(a). The magnetic domains were found to have average widths of $(390\pm7)\,\rm{nm}$ and $(554\pm11)\,\rm{nm}$ for the $500\,\rm{nm}$ and $1\,\rm{\mu m}$ thick films, respectively. This is in good agreement with Kittel's prediction that the domain width of a thick film with uniaxial anisotropy will scale proportionally to the square root of the thickness of the film \cite{Kittel46}. 

\begin{table}
 \begin{tabular}{ l  l  l  l  l } 
 \hline \hline
{\bf Thickness} & \bf{ X-ray energy}& $\mathbf{A_{\rm{XMCD}}}$ \bf{(a.u.)} &  {\bf FRC }& {\bf ES} \\ 
 \hline
 $1\,\rm{\mu m}$ & $7.2445\,\rm{keV}$ (Gd)  & 11 & $45\,\rm{nm}$ & $60\,\rm{nm}$\\ 
 \hline
 $500\,\rm{nm}$ &$7.2430\,\rm{keV}$ (Gd)  & 4 &  $118\,\rm{nm}$ & $120\,\rm{nm}$ \\ 
 \hline
 $1\,\rm{\mu m}$ &$7.1130\,\rm{keV}$ (Fe)   & 1 & $111\,\rm{nm}$ & $125\,\rm{nm}$\\ 
 \hline \hline
\end{tabular}
\caption{Comparison of the spatial resolution of different images calculated with Fourier ring correlation (FRC) 
and edge sharpness (ES) criteria. The strength of the $A_{\rm{XMCD}}$ signal of each image is normalized to 
the value of the $A_{\rm{XMCD}}$ signal at the Fe K edge for ease of comparison.}
\label{table:kysymys}
\end{table}


The effect of the strength of the XMCD signal on the spatial resolution was determined by comparing the spatial resolution of images of the $500\,\rm{nm}$-thick film measured at the Gd $L_{\rm{3}}$ edge (Figure \ref{fig:spat_res} (a)) and the $1\,\rm{\mu m}$ film measured at the Fe $K$ edge (Figure \ref{fig:spat_res} (b)) with the $45\,\rm{nm}$ spatial resolution that was measured for the $1\,\rm{\mu m}$-thick film at the Gd $L_{\rm{3}}$ edge. The relative strengths of the $A_{\rm{XMCD}}$ signal are shown in Table \ref{table:kysymys}. The spatial resolution was determined using a Fourier ring correlation (FRC) method and a profile across the domain boundaries, which was an average of five neighboring line profiles corresponding to a wall length of $200\,\rm{nm}$. The Fourier ring correlation curves are shown in more detail in Figure \ref{fig:spat_res}(c). It can be seen that the FRC of the image measured at the Fe $K$ edge (red line) is much lower than that of the $500\,\rm{nm}$-thick film (blue line) throughout the frequency range, which reflects the lower signal-to-noise ratio of the image of the $1\,\rm{\mu m}$ film measured at the Fe $K$ edge. Even though there is a large difference in signal-to-noise ratios, the images were found to have similar spatial resolutions, with the values for each method given in Table \ref{table:kysymys}. In particular, for the weaker contrast images, the edge sharpness measured with the line profile was reduced to $120\,\rm{nm}$ and $125\,\rm{nm}$ for the thinner film and weaker signal at the Fe $K$ edge, respectively, which agrees with the Fourier ring correlation measurements to within $10\%$.

In summary, we have achieved a spatial resolution of $45\,\rm{nm}$ in hard X-ray magnetic imaging by means of ptychography. This is a significant advance in the capabilities of X-ray magnetic imaging and will offer access to the magnetic configurations of a wide range of samples that up to now would have been impossible to measure.{
The measurement of high spatial resolution XMCD images at many different sample orientations combined with an appropriate tomographic reconstruction technique will make possible the mapping of the magnetization vector field within micrometer-size magnetic systems with nanoscale resolution. With first examples of magnetic tomography demonstrated with soft X-ray \cite{Streubel15}, neutron \cite{Kardjilov08} and electron microscopy \cite{Phatak14}, this work is therefore a first step towards applying the advantages of hard X-rays for tomographic imaging \cite{Holler14} of magnetic systems. } Moreover, with future brighter diffraction limited X-ray sources offering an expected increase in coherent flux of 2-3 orders of magnitude \cite{Eriksson14,Thibault14}, hard X-ray dichroic ptychography has the potential to provide single digit nanometer resolution and be extended towards the investigation of thin film heterostructures or architectures.

\section*{Acknowledgments}
The authors would like to thank Elisabeth M{\"u}ller and Michael Horisberger for help with sample fabrication, Patrick Ferrand, Marianne Liebi, Ben Watts, Laurence Bouchenoire and Olga Safonova for fruitful discussions. Hard X-ray dichroic ptychography experiments were performed at the cSAXS beamline, Swiss Light Source (SLS), Paul Scherrer Institut, Switzerland, XMCD fluorescence measurements were performed at the ID-12 beamline, European Synchrotron Radiation Facility (ESRF), France, and preliminary X-ray absorption spectroscopy measurements were performed at the SuperXAS beamline, Swiss Light Source (SLS), Paul Scherrer Institut, Switzerland.

\appendix

\section{Sample Characterization}\label{materials_and_methods}
\begin{figure}
  \centering
\includegraphics[width=0.5\textwidth]{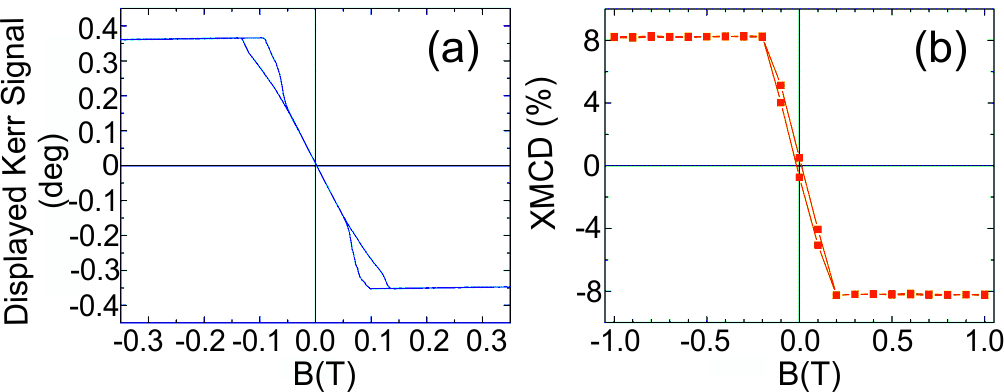}
	\caption{(a) Out of plane hysteresis loop of the $1\,\rm{\mu m}$ FeGd multilayer film measured by MOKE in the polar arrangement. (b) The XMCD signal of the reference $1\,\mu m$-thick FeGd multilayer film as a function of applied magnetic field, measured by X-ray fluorescence.}
\label{fig:supp_1}
\end{figure}

For image alignment purposes, platinum structures were patterned on the film. These were deposited with focused electron beam deposition instead of focused ion beam in order to avoid damaging the membrane. 
The magnetic properties of the FeGd films were characterized with magneto-optical Kerr effect (MOKE) measurements in the polar geometry. The hysteresis loop of the ${\rm 1\, \mu m}$ thick film is given in Figure \ref{fig:supp_1}(a). The film has perpendicular magnetic anisotropy with a saturating field of $130\,{\rm mT}$. The hysteresis at approximately $100\,{\rm mT}$, together with zero remanence at zero field, indicates the spontaneous nucleation and formation of magnetic domains \cite{Suszka14,Tripathi11}. 

A similarly prepared sample was measured with XMCD fluorescence as a reference for the XMCD results measured with ptychography. To characterize the magnetic properties of the sample, the out-of-plane magnetic hysteresis was investigated by measuring the XMCD signal with X-ray fluorescence as a function of magnetic field applied perpendicular to the sample surface. The hysteresis and non-zero-magnetization at remanence indicate that the sample also has an out-of-plane magnetization component.

\section{XMCD Fluorescence Measurements}\label{fluorescence}

X-ray absorption spectroscopy experiments were performed at the  ID12 beamline of the European Synchrotron Radiation Facility (ESRF), France at the $L_{\rm 3}$ edge of Gd and at the $K$ edge of Fe using a highly efficient fluorescence yield detection mode in a backscattering geometry \cite{Rogalev01}. For the XMCD measurements, circularly polarized X-rays generated by a helical undulator of the APPLE-II type were used. At these photon energies the circularly polarized beam emitted by the helical undulator is reduced to $91-92\%$, due to the partial suppression of horizontal polarization at the Bragg angles of the Si $\textless {111} \textgreater$ monochromator. The spectra were corrected for the incomplete circular polarization. 

The sample was inserted into the bore of a superconducting magnet. The measurements were carried out at room temperature and under a magnetic field of $1\,{\rm T}$ applied perpendicular to the sample surface.

XMCD spectra were obtained as a difference of X-ray absorption spectra recorded with right and left circularly polarized X-rays. To ensure that the obtained XMCD spectra were free of any experimental artefacts, the data were collected for both directions of the external magnetic field applied parallel and antiparallel to the incoming X-ray beam and the XMCD measurements compared and found to be equivalent.

A reference spectrum for the phase XMCD was calculated using the Kramers-Kronig relations \cite{Watts14} assuming the atomic density of bulk gadolinium ($2.38\times10^{28}\,{\rm atoms/m^3}$) and a nominal effective thickness of ${\rm 520\, nm}$. This was compared with the phase part of the ptychographic XMCD data, $\phi_{XMCD}$. 

\section{Ptychography experimental setup}\label{setup}

X-ray ptychographic imaging was performed at the cSAXS beamline at the Swiss Light Source (SLS), Switzerland with X-rays produced by an in-vacuum undulator with $19\,{\rm mm}$ period, operated at a gap of ${\rm 5.338\, mm}$. The experiment was performed in air and at room temperature.

The illumination on the sample was defined by a combination of a $35\,{\rm \mu m}$ central stop, a gold Fresnel zone plate with $1.2\,{\rm \mu m}$ gold thickness, $60\,\rm{ nm}$ outermost zone width, and $170\,{\rm \mu m}$ diameter, and a $20\,{\rm \mu m}$ diameter order-sorting aperture. The X-ray beam incident on the sample had a coherent flux of approximately $3\times 10^7$ photons/s and a spot size of $4.5\,{\rm \mu m}$. A  diffraction pattern was measured at each of the scanning points with a 0.2 second exposure time using a single photon counting Pilatus 2M detector \cite{Henrich09,Kraft09}. To reduce air scattering and absorption, a helium filled flight tube was placed between the sample and detector. 
Measurements were taken at sixteen energies between $7.234\,\rm{ keV}$ and $7.2565\,\rm{keV}$ for the Gd $L_{\rm 3}$ edge measurements. To enhance the signal-to-noise ratio, ten such scans were repeated for each energy. These scans were measured with left and right circularly polarized light. 

For the Fe K edge measurements, ptychography was performed with $\rm{C_L}$ polarized X-rays only due to time constraints.  At ten energies between $7.109\,\rm{keV}$ and $7.118\,\rm{keV}$, sixty absorption images were measured and averaged. The quantitative XMCD values were obtained by fitting the histograms of the absorption XMCD contrast of the pixels of the positive and negative magnetic domains to Gaussian distributions and calculating the mean and standard deviation. 

\section{Ptychographic reconstructions}\label{recons}

Ptychographic reconstructions were performed with the data from $192 \times 192$ pixels of the Pilatus detector, resulting in a reconstruction pixel size of approximately 37 nm and 38 nm for measurements taken at the Gd $L_{\rm{3}}$ edge and Fe K edge, respectively. 

The algorithm initial guess was given by a uniform transmission for the object and an illumination obtained through a previous ptychography measurement on a strongly scattering test object. The initial guess for the illumination was numerically propagated for each different energy in the series to account for the dependence of the Fresnel zone plate focal distance on the photon energy. 

The reconstructed images were aligned to a small fraction of a pixel \cite{Guizar08} and a linear phase ramp was removed from the images as in Ref.\onlinecite{Guizar11} before the images were averaged. 

\begin{figure}
  \centering
\includegraphics[width=0.48\textwidth]{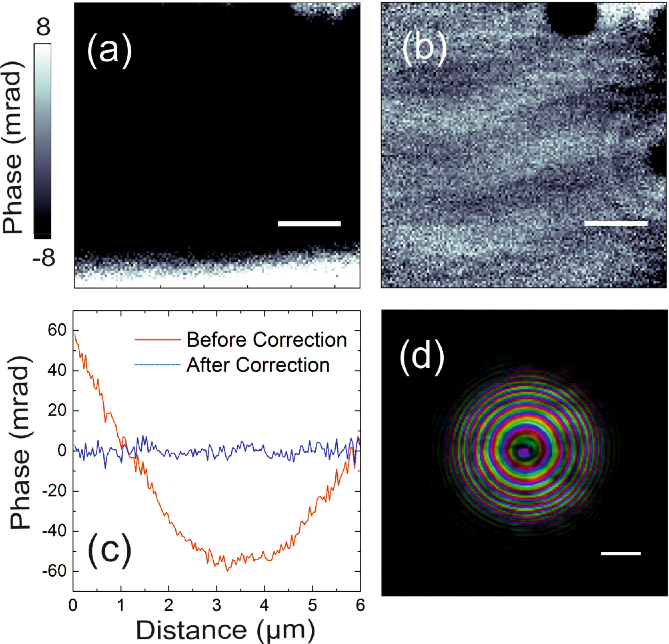}%
	\caption{(a) The smooth phase artifact obscurs the fine detail of the image, which is then subtracted to reveal both magnetic and electronic contrast in (b). A line profile taken vertically, both before and after removal of the artefact, is shown in (c), with the background becoming relatively flat after correction. (d) The reconstructed illumination at a photon energy of 7.2430 keV for a scan of the magnetic domains. The amplitude is encoded in the brightness of the image, whilst the phase of the probe is shown as the hue. Scale bars represent $1\,\rm{\mu m}$.}
\label{fig:supp_4}
\end{figure}

A smoothly spatially varying phase artifact with an amplitude of a few milliradians was present in the reconstructions. The artifacts varied between different measurements and it is believed that they may arise in the reconstruction as a result of an algorithmic compensation mechanism for small changes in the probe wavefront. In order to correct for this artifact, the phase image was projected onto a two dimensional polynomial orthogonal basis using least-squares fitting similar to that performed in Ref. \onlinecite{Lima13}. An example of this artifact and its subsequent removal is shown in Figure \ref{fig:supp_4} for a single reconstruction.

After normalization, ten images for each energy and polarization were averaged, and the ${\rm A_{XMCD}}$ and ${\rm \phi_{XMCD}}$ images were calculated. As discussed in the main text, ptychography assumes a scalar product interaction as follows:
\begin{equation}
\mathbf{\Psi}_j(\mathbf{r}) = \mathbf{P(r)O(r-r}_j\mathbf{)}
\end{equation}
Where $\mathbf{\Psi}_j(\mathbf{r})$ is the exit field after the sample, $\mathbf{P(r)}$ is the incident illumination, $\mathbf{O(r-r}_j\mathbf{ )}$ is the transmissivity of the sample, all complex-valued functions of $\mathbf{r}$, the transverse Cartesian coordinates, and $j$ indicates the ptychography scan position.There is a certain amount of ambiguity when separating the reconstructed probe and object, and the absolute value of the reconstruction is subject to a multiplicative factor $D$ as follows:
\begin{equation}
\mathbf{\Psi}_j(\mathbf{r}) = \frac{\mathbf{P(r)}}{D}\bigg(\mathbf{O(r-r}_j)\cdot D\bigg)
\end{equation}
which translates to a multiplicative factor for the absorption part, and an additive offset to the phase part of the reconstruction. To obtain the quantitative absorption and absolute phase values, the reconstructions were normalized by removing a phase offset and a multiplicative factor for the phase and absorption parts of the reconstructions, respectively. For a ${\rm \phi_{XMCD}}$ image, the average phase of the reconstruction of either helicity was calculated by taking the mean phase value of an equal number of positive and negative domains, and the net offset between the two helicities removed to leave purely magnetic contrast. A value for ${\rm \Delta_{edge}}$ was found from a reference transmission spectrum measured on a similar sample and used to normalize the absorption data. For a sample with an area of air within the image, this ambiguity is removed by using the air as a reference for normalization. 

\section{Estimate of the XMCD Contrast}

\begin{figure}
  \centering
\includegraphics[width=0.5\textwidth]{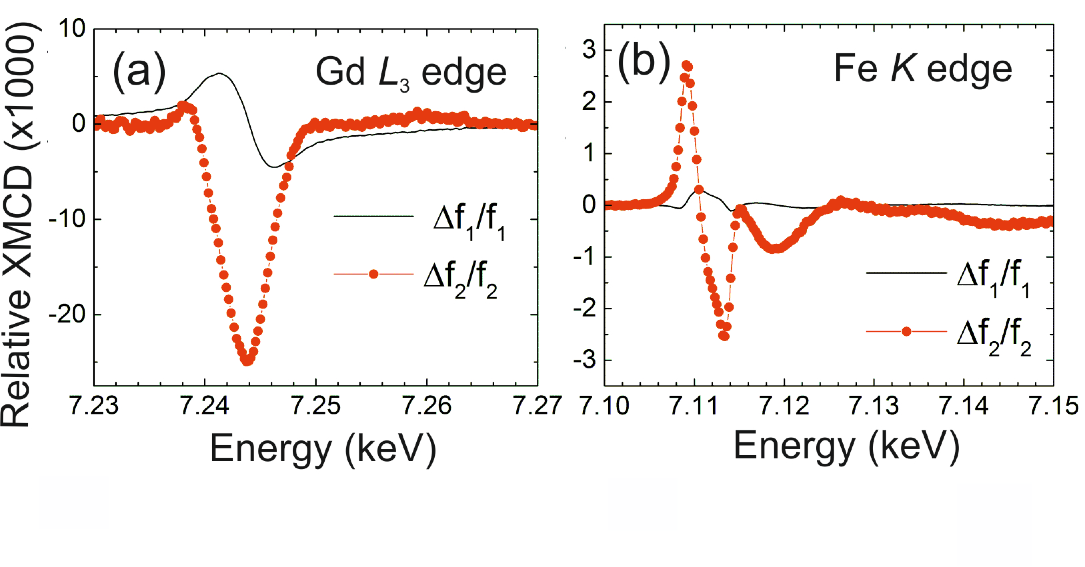}
	\caption{Comparison of the relative phase ($\Delta f_1/f_1$, black line) and absorption ($\Delta f_2/f_2$, red dots) magnetic contrast (a) across the Gd $L_{\rm 3}$ edge and (b) across the Fe $K$ edge. At the Fe K edge it can be seen that  the expected phase contrast is much smaller than the absorption contrast.}
\label{fig:supp_6}
\end{figure}

The expected values for the relative XMCD contrast are given in terms of the real ($f_1$) and imaginary ($f_2$) parts of the scattering factor as $\Delta f_1/f_1$ (black line) and $\Delta f_2 / f_2$ (red dots) for  the Gd $L_3$ edge and the Fe $K$ edge in Figures \ref{fig:supp_6}(a) and \ref{fig:supp_6}(b), respectively. The real part of the scattering factor, $f_1$, is calculated with the Kramers-Kronig relations from the spectra measured in fluorescence ($f_2$). This allows the strength of the phase ($\Delta f_1 / f_1$) and absorption ($\Delta f_2 / f_2$) magnetic contrast at the Gd $L_{\rm 3}$ and Fe $K$ edges to be compared. The expected absorption magnetic contrast, normalized by the total signal ($\Delta f_2 / f_2$), is approximately 10 times weaker at the Fe K edge than the Gd $L_{\rm 3}$ edge. However, the expected phase magnetic contrast ($\Delta f_1 / f_1$) at the Fe $K$ edge is almost 20 times weaker than at the Gd $L_{\rm 3}$ edge, with a contrast of approximately $0.1\,\rm{mrad}$ for the $1\,\rm{\mu m}$-thick FeGd film.  However, a layer of carbon was deposited on the surface of the film at the region of interest during the experiment due to the sample not being in a vacuum environment. As the phase signal of the carbon deposition ($ \geq 0.5\,{\rm mrad}$) was greater than the expected $\phi_{\rm XMCD}$ contrast, it was not possible to extract the $\phi_{\rm XMCD}$ contrast from the images at the Fe K edge in this experiment.

\end{document}